\documentclass[12pt]{iopart}

\usepackage[]{cite}

\usepackage[]{amssymb}
\usepackage[]{mathrsfs}
\usepackage[]{eucal}
\usepackage[]{tensor}
\usepackage[]{amsthm}
\usepackage[]{bm}

\newcommand{\dd}{\partial}
\newcommand{\df}{\mathrm{d}}

\newcommand{\Lie}{\pounds}
\newcommand{\nab}[1]{\nabla_{\!#1}}

\newcommand{\bc}{\begin{center}}
\newcommand{\ec}{\end{center}}
\newcommand{\be}{\begin{equation}}
\newcommand{\ee}{\end{equation}}

\usepackage{dsfont}

\newcommand{\rr}{\mathds{R}}

\newcommand{\0}{\emptyset}

\usepackage[unicode]{hyperref}
\usepackage{xcolor}
\definecolor{pastgreen}{HTML}{669900}
\definecolor{pastblue}{HTML}{336699}
\definecolor{linkcol}{HTML}{663333}
\hypersetup{colorlinks,linkcolor={pastblue},citecolor={pastblue},urlcolor={pastblue}}

\theoremstyle{plain} \newtheorem{tm}{Theorem}[section]
\theoremstyle{plain} \newtheorem{lm}[tm]{Lemma}
\theoremstyle{plain} \newtheorem{cor}[tm]{Corollary}
\theoremstyle{definition} \newtheorem{defn}[tm]{Definition}
\theoremstyle{definition} 
\newcommand{\btm}{\begin{tm}}
\newcommand{\etm}{\end{tm}}
\newcommand{\blm}{\begin{lm}}
\newcommand{\elm}{\end{lm}}
\newcommand{\bcor}{\begin{cor}}
\newcommand{\ecor}{\end{cor}}
\newcommand{\bdefn}{\begin{defn}}
\newcommand{\edefn}{\end{defn}}

\begin{document}

\begin{flushright}
ZTF-EP-17-08
\end{flushright}

\title{On symmetry inheritance of nonminimally coupled scalar fields}

\author{Irena Barja\v si\'c and Ivica Smoli\'c}
\address{Department of Physics, Faculty of Science, University of Zagreb, Bijeni\v cka cesta 32, 10000 Zagreb, Croatia}
\eads{\mailto{irena.barjasic@hotmail.com}, \mailto{ismolic@phy.hr}}

\date{\today}

\begin{abstract}
We present the first symmetry inheritance analysis of fields non\-mi\-ni\-mally coupled to gravity. In this work we are focused on the real scalar field $\phi$ with nonminimal coupling of the form $\xi\phi^2 R$. Possible cases of the symmetry noninheriting fields are constrained by the properties of the Ricci tensor and the scalar potential. Examples of such spacetimes can be found among those which are ``dressed'' with the stealth scalar field, a nontrivial scalar field configuration with the vanishing energy-momentum tensor. We classify the scalar field potentials which allow the symmetry noninheriting stealth field configurations on top of the exact solutions of the Einstein's gravitational field equation with the cosmological constant.
\end{abstract}

\pacs{04.20.Cv, 04.20.Jb, 04.40.-b, 04.70.Bw}

\vspace{2pc}

\noindent{\it Keywords}: symmetry inheritance, nonminimally coupled scalar fields, stealth fields, no-hair theorems


\section{Introduction}

Just as in Newtonian gravity, the isometries of the spacetime and the symmetries of the matter fields have proven to be an invaluable assumption in numerous analyses of the general relativity. One of the basic questions one encounters in the construction of a symmetric spacetime is whether the matter fields \emph{inherit} the spacetime symmetries. The earliest works on this topic have appeared in the mid-1970s \cite{Woo73a,Woo73b,MW75,Coll75,RT75,WY76a,WY76b,Hoen78}, after which we have had only occasional bursts of activity, separated by relatively dormant phases. The most recent series of papers \cite{ISm15,CDPS16,ISm17,BGS17} have solved some long standing open problems and opened several new ones. However, all the results on the symmetry inheritance, both in the old as well as in the more recent papers, are about matter fields \emph{minimally coupled} to gravity. As all the previously used techniques have proven to be inadequate in this context \cite{ISm15}, the attack on the nonminimally coupled fields requires some new ideas. In this paper we shall be mainly focused on real scalar fields, nonminimally coupled to gravity. Lagrangian terms with nonminimally coupled scalar fields are generated by quantum corrections \cite{BD} and are in fact necessary for the theory to be renormalizable in a classical gravitational background \cite{CCJ70}. Models with nonminimally coupled scalar fields have proven to be interesting for the inflatory cosmological models \cite{FM87,Faraoni96}, as well as the important source of the black hole hair \cite{Saa96a,Saa96b,AB02,Winst05,HW09}.

\medskip

Symmetry inheritance belongs to those self-evident assumptions that are often (and too easily) taken for granted. For example, symmetry inheritance of scalar fields is assumed in Bekenstein's no-hair theorems \cite{Beken72,BekenStaticHair,BekenStationaryHair} and their generalizations \cite{XZ91,Beken95,Sud95,Zann95,Saa96a,Saa96b,SZ98,AB02,Winst05,BL07,BL11,SS14,Hod17}, as well as in the staticity and the circularity theorems (see e.g. \cite{Heusler}, section 12.1). Better understanding of the possible forms of the black hole solutions is especially important since we are entering an era of intensive observations of gravitational phenomena, both in the electromagnetic spectrum \cite{BJLP14,CHRR15,Test16,CG16,J16,Cast17} as well as with gravitational waves \cite{BCW06,GVS12,Berti15,CG16,LIGO16,CFP16,CHMPP16,CFMPR17,CP17,BCRG17}, which will allow us precise testing of the various hairy black hole solutions. Our aim here is to shed some light on general symmetry inheriting properties of the nonminimally coupled scalar fields and to pave a systematic approach to the problem of the existence of the symmetry noninheriting real scalar black hole hair.

\medskip

The paper is organized as follows. In section 2 we explain the relevant details of the field equations. In section 3 we introduce several strategies of the symmetry inheritance analysis for the nonminimally coupled scalar fields and then apply these ideas in section 4 to various cases, with different scalar potentials and different physical parameters of the model. In section 5 we discuss the relation of the symmetry inheritance constraints with the no-hair theorems. In section 6 we give a brief overview of the stealth scalar fields, which are the first obvious candidates for the nonminimally coupled symmetry noninheriting fields, and present a classification of the scalar potentials which admit such solutions. In the last section we give some final remarks and point out remaining open questions.

\medskip

Throughout the paper we follow the choices of conventions from \cite{Wald}, the ``mostly plus'' metric signature and the natural system of units, $G = c = 1$. Unless stated otherwise, the total number of spacetime dimensions is always general $D \ge 3$ and all the fields and functions are assumed to be sufficiently smooth.

\medskip

\section{Field equations}

We shall consider an action which is a sum of three parts, the gravitational $S_g$, the scalar $S_\phi$ and the mixed one $S_{\phi g}$, 
\be
S = S_g + S_\phi + S_{\phi g} = \int \df^D x \, \sqrt{-g} \, \left( \frac{1}{16\pi}\,L_g + L_\phi + L_{\phi g} \right) \ .
\ee
The contribution of the gravitational part to the field equations, 
\be
E_{ab} = \frac{16\pi}{\sqrt{-g}}\,\frac{\delta S_g}{\delta g^{ab}}
\ee
is assumed to be a general diff-covariant smooth function of the spacetime metric, the Riemann tensor, the Levi-Civita tensor and the covariant derivatives. The scalar field appears in the remaining two parts of the action,

\begin{itemize}
\item[(i)] the canonical part,
\be
L_\phi = X - V(\phi) \ ,
\ee
consisting of the kinetic term
\be
X \equiv -\frac{1}{2}\,\nab{c}\phi\,\nabla^c\phi
\ee
and the potential $V(\phi)$;

\item[(ii)] the mixed gravitational-scalar sector which represents the nonminimal coupling,
\be
L_{\phi g} = -f(\phi)R \ .
\ee
\end{itemize}

\smallskip

Although the large part of the discussion below may be applied to a general smooth function $f$, our interest is mainly on the choice
\be\label{eq:fxiphi2}
f(\phi) = \frac{1}{2}\,\xi\phi^2
\ee
with some real parameter $\xi \ne 0$. There is a special choice of the parameter $\xi$,
\be
\xi_c = \frac{D-2}{4(D-1)} \ ,
\ee
which leads to conformally invariant field equations (see e.g.~\cite{Wald}, Appendix D). The value of the parameter $\xi$ has been recently constrained \cite{Hryc17} by the cosmological observations, using the flat FRW model and the nonminimally coupled scalar field with vanishing potential $V = 0$. 

\medskip

The corresponding energy-momentum tensor $T_{ab}^{(\phi)}$ is introduced via
\be
-\frac{2}{\sqrt{-g}}\,\frac{\delta (S_\phi + S_{\phi g})}{\delta g^{ab}} = T_{ab}^{(\phi)} + 2 f(\phi) G_{ab} \ ,
\ee
so that
\be
T_{ab}^{(\phi)} = \nab{a}\phi\nab{b}\phi - 2\nab{a}\nab{b} f(\phi) + \Big( X - V(\phi) + 2\,\Box f(\phi) \Big) g_{ab} \ .
\ee
Furthermore, using the decomposition
\be
\nab{a}\nab{b} f(\phi) = f''(\phi)\nab{a}\phi\nab{b}\phi + f'(\phi)\nab{a}\nab{b} \phi
\ee
we can write the energy-momentum in the following form
\begin{eqnarray}
T_{ab}^{(\phi)} & = \big( 1 - 2f''(\phi) \big) \nab{a}\phi\nab{b}\phi - 2f'(\phi)\,\nab{a}\nab{b}\phi \, + \nonumber\\
 & + \Big( (1-4f''(\phi))X - V(\phi) + 2f'(\phi)\,\Box\phi \Big) g_{ab} \ . \label{eq:Tab}
\end{eqnarray}
Here we also introduce the trace of the energy-mo\-men\-tum tensor,
\begin{eqnarray}
T \equiv g^{ab} T_{ab}^{(\phi)} & = -4(D-1)(f''(\phi) - \xi_c) X - \nonumber\\
 & - DV(\phi) + 2(D-1)f'(\phi)\,\Box\phi \ .
\end{eqnarray}
Finally, the complete set of field equation in our problem consists of the gravitational one,
\be\label{eq:EOM}
- 16\pi f(\phi) G_{ab} + E_{ab} = 8\pi T_{ab}^{(\phi)} \ ,
\ee
where $G_{ab}$ is the Einstein's tensor, and the generalized Klein-Gordon equation,
\be\label{eq:KG}
\Box\phi = V'(\phi) + f'(\phi)R \ .
\ee

\medskip

\section{Strategies of analysis}

The object of our discussion is a smooth Lorentzian manifold $(M,g_{ab},\phi)$ with a smooth scalar field $\phi$, admitting a smooth Killing vector field $K^a$. We are interested under which assumptions the open set
\be\label{eq:setS}
S = \left\{ \, p \in M \, : \ \Lie_K \phi(p) \ne 0 \, \right\}
\ee
is \emph{empty}, in which case we will say that the field $\phi$ inherits the symmetry. Henceforth we focus our analysis on the points of $S$. For later convenience we also introduce a closed set $Z$ of zeros of the field $\phi$,
\be\label{eq:setZ}
Z = \left\{ \, p \in M \, : \ \phi(p) = 0 \, \right\} \ .
\ee
Note that the interior of a set where $\phi$ has some constant value, $\phi = \phi_0$, is a subset of $M - S$. Thus, for example, $Z^\circ \cap S = \0$. Also, as at any zero of the Killing vector field, a point where $K^a = 0$, we have $\Lie_K \phi = K^a \nab{a} \phi = 0$, these points are never elements of the set $S$.

\medskip

As a consequence of the assumptions about the tensor $E_{ab}$ we know that the existence of a Killing vector field $K^a$, such that $\Lie_K g_{ab} = 0$, implies $\Lie_K E_{ab} = 0$. For a \emph{minimally} coupled field this would immediately lead to the condition $\Lie_K T_{ab} = 0$, a crucial equation in the recent detailed analyses of the symmetry inheritance for real and complex scalar fields \cite{ISm15,ISm17}, as well as the electromagnetic field \cite{CDPS16,BGS17}. However, as we are looking at nonminimally coupled fields, we have to resort to other strategies. Note that it is possible to transform nonminimally coupled to minimally coupled scalar field via conformal transformation of the form $\bar{g}_{ab} = \Omega^{-2}(\phi) g_{ab}$ with some appropriate real function $\Omega$ \cite{Saa96a}. But, as was already remarked in \cite{ISm15}, unless we know \emph{a priori} that $\Lie_K \phi = 0$, there is no guarantee that $\Lie_K g_{ab} = 0$ will imply $\Lie_K \bar{g}_{ab} = 0$, so that this trick is not useful for the analysis of the symmetry inheritance.

\subsection{The conformal case}

Let us first look at the distinguished case of the conformal coupling, $f(\phi) = \xi_c \phi^2/2$. If we act on both sides of the trace of the gravitational field equation (\ref{eq:EOM}),
\be\label{eq:traceEOM}
g^{ab} E_{ab} + 8\pi \Big( DV(\phi) - \frac{D-2}{2}\,\phi V'(\phi) \Big) = 0 \ ,
\ee
with the Lie derivative $\Lie_K$, we get 
\be
W_c(\phi) \Lie_K \phi = 0 \ ,
\ee
where we have introduced an auxiliary function
\be\label{eq:Wc}
W_c(\phi) \equiv (D+2)V'(\phi) - (D-2)\phi V''(\phi) \ .
\ee
This implies the following result.

\medskip

\btm\label{tm:confxi}
Let $(M,g_{ab},\phi)$ be a solution to the field equations (\ref{eq:EOM})--(\ref{eq:KG}) with $f(\phi) = \xi_c \phi^2/2$ and $D \ge 3$, admitting a smooth Killing vector field $K^a$. Then the support of the function $W_c$, defined in (\ref{eq:Wc}), is disjoint from the set $S$, defined in (\ref{eq:setS}).
\etm

\medskip

In other words, for each point such that $W_c(\phi) \ne 0$, the symmetry is necessarily inherited, $\Lie_K \phi = 0$. For example, if $V = V_{\mathrm{mass}} = \mu^2 \phi^2/2$ we have $W_c(\phi) = 4\mu^2\phi$, thus in the massive case ($\mu \ne 0$) the set $S - Z$ is empty. We could ask for which potential $V(\phi)$ the function $W_c(\phi)$ \emph{identically} vanishes (i.e.~it is zero for \emph{any} real scalar field $\phi$ on the domain of the function $W_c$). The solution of the ordinary differential equation $W_c = 0$ for the unknown $V$ is given by
\be\label{eq:Vc}
V_c(\phi) = A \phi^{\frac{2D}{D-2}} + B \ ,
\ee
where $A$ and $B$ are some real constants. The theorem \ref{tm:confxi} does not tell us anything about the set $S$ for this particular choice of the potential and, as will be explained in section 5, here one can find the examples of symmetry noninheriting scalar fields. We can, however, get one useful constraint in this special subcase: By inserting the potential (\ref{eq:Vc}) back in the equation (\ref{eq:traceEOM}), it follows that the trace of the tensor $E_{ab}$ is constant, 
\be\label{eq:traceE}
g^{ab} E_{ab} = -8\pi DB \ .
\ee
Note that the constant $B$ effectively only contributes to the cosmological constant $\Lambda$, thus without any loss of generality, we can set $B = 0$, which reduces the constraint (\ref{eq:traceE}) to the tracelessness of the tensor $E_{ab}$. For other choices of the nonminimal coupling the trace of the gravitational field equation will not be as informative as in the conformal case, thus we have to find yet another way to attack the problem.

\subsection{A strategy aided by Frobenius}

Let us now suppose, in addition to the previous assumptions, that the Killing vector field $K^a$ is hypersurface orthogonal, so that it satisfies the Frobenius' condition
\be
K_{[a} \nab{b} K_{c]} = 0 \ .
\ee
It is well known \cite{Heusler} that this implies the Ricci staticity condition,
\be\label{eq:Riccistat}
K_{[a} R_{b]c} K^c = 0 \ .
\ee
However, there is even a larger class of symmetric tensors $E_{ab}$, satisfying the analogous property 
\be\label{eq:ot1}
K_{[a} E_{b]c} K^c = 0 \ .
\ee
Such tensors were, in somewhat broader context, labeled as the members of the \emph{orthogonal-transitive class} (of order 1) in \cite{ISm17}. Beside the canonical Einstein's tensor with the cosmological constant term, we know that the Lovelock tensor \cite{Love71} and the tensor derived from $f(R)$ theories \cite{NO06,NO10,SF10,DeFT10} also satisfy the condition (\ref{eq:ot1}). As our method of analysis is applicable to any such a tensor, we shall now assume that the tensor $E_{ab}$ in the field equation (\ref{eq:EOM}) belongs to the orthogonal-transitive class.

\medskip

Each point $p \in M$ where $K^a \ne 0$ has a coordinate chart $(O_p,\psi)$ on which one of the components of the Killing vector field, say $K^1$, is nonvanishing. Thus, we can always find an auxiliary (smooth) vector field $L^a$, such that $L^a K_a \ne 0$ on this neighbourhood. Then, from the contraction of the equation (\ref{eq:ot1}) with $L^a$, it follows that on a neighbourhood of any point where $K^a \ne 0$ we have
\be
E_{ab} K^b = \gamma_{\mathrm{E}} K_a
\ee
for some real function $\gamma_{\mathrm{E}}$, such that
\be
\Lie_K \gamma_{\mathrm{E}} = 0 \ .
\ee
Likewise, there is a function $\gamma_{\mathrm{G}}$ such that
\be
G_{ab} K^b = \gamma_{\mathrm{G}} K_a
\ee
for the Einstein's tensor $G_{ab}$, and $\Lie_K \gamma_{\mathrm{G}} = 0$.

\medskip

The first step is to extract the $X$ from the trace of the field equation (\ref{eq:EOM}), which can be done at all points where $f''(\phi) \ne \xi_c$. We have
\be\label{eq:XT}
X = \frac{T + DV(\phi) - 2(D-1)f'(\phi)\,\Box\phi}{-4(D-1)(f''(\phi) - \xi_c)} \ ,
\ee
where $T$ is to be replaced by the trace of the lhs of the field equation (\ref{eq:EOM}) and the D'Alembertian $\Box\phi$ with the rhs of the Klein-Gordon equation (\ref{eq:KG}). Next we contract the field equation (\ref{eq:EOM}) with $K^a\nabla^b\phi$,
\be\label{eq:EOMKnabphi}
\left( -2\gamma_G f(\phi) + \frac{1}{8\pi}\,\gamma_{\mathrm{E}} \right) \Lie_K \phi = T_{ab}^{(\phi)} K^a \nabla^b \phi \ .
\ee
The main idea now is to rewrite the rhs in the same form as the lhs, namely, as the Lie derivative $\Lie_K \phi$ multiplied by some algebraic expression. This can be achieved with the help of the identity
\be
K^a (\nabla^b \phi) \, \nab{a} \nab{b} \phi = -\Lie_K X
\ee
and the relation (\ref{eq:XT}). At the end one can obtain an algebraic relation that must be satisfied on the set $S$. In order to simplify the expressions and reach some more concrete conclusions in the rest of the discussion we shall assume that the function $f(\phi)$ is given by (\ref{eq:fxiphi2}).

\medskip

\section{Constraining the symmetry inheritance}

Let us begin with the simplest case, the Einstein's field equations ($E_{ab} = G_{ab}$) and the vanishing scalar potential $V(\phi) = 0$. First, if $\xi \ne \xi_c$, we get that
\be
X = \left( \frac{1}{2}\,\xi\phi^2 + \frac{\xi_c}{16\pi(\xi - \xi_c)} \right) R
\ee
and
\be
\Lie_K X = \xi R \phi \Lie_K \phi \ .
\ee
This allows us to compute the crucial contraction of the energy-momentum tensor $T_{ab}^{(\phi)} K^a \nabla^b \phi$. At any point of the set $S$ the equation (\ref{eq:EOMKnabphi}) implies the relation
\be\label{eq:gammaG}
\left( -\xi\phi^2 + \frac{1}{8\pi} \right) \gamma_G = \left( \frac{1}{2}\,(8\xi - 1) \xi\phi^2 - \frac{\xi_c}{16\pi(\xi - \xi_c)} \right) R \ .
\ee
Taking the Lie derivative $\Lie_K$ of this relation gives us
\be
\Big( 2\xi\gamma_G + (8\xi - 1)\xi R \Big) \phi \Lie_K \phi = 0 \ .
\ee
So, at each point of the set $S - Z$, we have
\be
\gamma_G = \frac{1}{2}\,(1 - 8\xi)R \ .
\ee
By inserting this back into (\ref{eq:gammaG}) we get a simple constraint
\be
\xi(\xi - \xi_*)R = 0 \ ,
\ee
where we have introduced an auxiliary parameter
\be
\xi_* \equiv \xi_c + \frac{1}{8} = \frac{3D-5}{8(D-1)} \ .
\ee
Thus if $\xi \notin \{0,\xi_c,\xi_*\}$ then we can conclude that $R = 0$ at all points of the set $S - Z$. Since the interior of $Z$ is disjoint from the set $S$, it is only $\dd Z \cap S$ that may be a nonempty subset of $Z \cap S$. But, using lemma 2 (which implies that equality of continuous maps on some set extends to the closure of that set) and lemma 3 (which implies that the closure of complement of the boundary of an open or a closed set is the whole topological space) from \cite{CDPS16}, we can in fact conclude that $R = 0$ at all points of the set $S$.

\medskip

What about the conformal case, $\xi = \xi_c$? Namely, the vanishing potential $V = 0$ is the $A = B = 0$ case of (\ref{eq:Vc}), for which the theorem \ref{tm:confxi} remains silent. The trace of the gravitational field equation, with $E_{ab} = G_{ab}$ and $V = 0$, gives
\be
\left( \frac{1}{8\pi}\,\xi_c + \xi(\xi - \xi_c)\phi^2 \right) R = 2(\xi - \xi_c)X \ .
\ee
So, in the conformal case $\xi = \xi_c$ we get again $R = 0$ (unless $\xi_c = 0$, which happens for $D = 2$). Unfortunately, in this case we cannot say anything useful about the kinetic term $X$. We can now summarize these conclusions in the following theorem.

\medskip

\btm\label{tm:zeroV}
Let $(M,g_{ab},\phi)$ be a smooth solution to the field equations (\ref{eq:EOM})--(\ref{eq:KG}) with $E_{ab} = G_{ab}$, $V(\phi) = 0$, $f(\phi) = \xi \phi^2/2$ and $\xi \notin \{0,\xi_*\}$, admitting a hypersurface orthogonal smooth Killing vector field $K^a$. Then $R = 0$ at all points of the set $S$, defined in (\ref{eq:setS}). If in addition $\xi \ne \xi_c$, then also $X = 0$ at all points of the set $S$.
\etm

\medskip

An immediate corollary is that, under the conditions given in the theorem above, if $R \ne 0$ at all points of the spacetime, then the set $S$ is empty and the scalar field $\phi$ necessarily inherits the symmetry. On the other hand, for $\xi \notin \{0,\xi_c,\xi_*\}$ the theorem \ref{tm:zeroV} guarantees that both the Ricci scalar $R$ and the kinetic term $X$ identically vanish on the set $S$, so that the field equations are reduced to the following system
\be
\Box\phi = 0 \ ,
\ee
\be\label{eq:redEOM}
\Big( \frac{1}{8\pi} - \xi\phi^2 \Big) R_{ab} = (1-2\xi) \nab{a}\phi \nab{b}\phi - 2\xi\phi\nab{a}\nab{b}\phi \ .
\ee
Let us now focus on the open set $S'$, defined as the intersection of the set $S$ with the (open) set of points where $\phi^2 \ne (8\pi\xi)^{-1}$. First, if we contract the gravitational field equation (\ref{eq:redEOM}) with $\nabla^b \phi$, at each point of the set $S'$ we get
\be
R_{ab} \nabla^{b}\phi = 0 ,
\ee
from which it follows that
\be
R_{ab} \nabla^a \nabla^b \phi = \nabla^b (R_{ab} \nabla^a \phi) - \frac{1}{2}\,\nabla^a \phi \nab{a} R = 0 \ .
\ee
The contraction of the equation (\ref{eq:redEOM}) with $R^{ab}$ then gives
\be\label{eq:RicciRicci}
\Big( \frac{1}{8\pi} - \xi\phi^2 \Big) R_{ab} R^{ab} = 0 \ .
\ee
Also, if we combine the equations (\ref{eq:Riccistat}) and (\ref{eq:redEOM}),
\be
(1 - 2\xi)(K_{[a}\nab{b]}\phi)\Lie_K \phi - 2\xi\phi K^c K_{[a}\nab{b]} \nab{c}\phi = 0 \ ,
\ee
then contract it with $K^a\nabla^b\phi$, on the set $S'-Z$ it implies the relation
\be
K^a K^b \nab{a}\nab{b}\phi = \frac{1-2\xi}{2\xi\phi}\,(\Lie_K \phi)^2 \ .
\ee
Using this in the equation (\ref{eq:redEOM}) contracted with $K^a K^b$ we get
\be\label{eq:RicciKK}
\Big( \frac{1}{8\pi} - \xi\phi^2 \Big) R_{ab} K^a K^b = 0 \ .
\ee
Equations (\ref{eq:RicciRicci}) and (\ref{eq:RicciKK}) provide us with two additional constraints: If $\xi \notin \{0,\xi_c,\xi_*\}$, at each point of the set $S'-Z$ we have $R_{ab} R^{ab} = 0$ and $R_{ab} K^a K^b = 0$. This suggests that the natural candidates for the symmetry noninheriting nonminimally coupled real scalar field with vanishing potential $V$ should be looked for among the Ricci-flat solutions.

\medskip

As a first step towards the generalization of this result, we can assume that the potential is given by the mass term $V = V_{\mathrm{mass}} \equiv \mu^2 \phi^2/2$, where $\mu \ge 0$, and add the nonvanishing cosmological constant, $E_{ab} = G_{ab} + \Lambda g_{ab}$. In this case, if $\xi \notin \{0,\xi_c\}$, at all points of the set $S - Z$ we have
\be
\gamma_G = \frac{1}{2}\,(1 - 8\xi)R - \frac{(4\xi - 1)^2}{4\xi(\xi - \xi_c)}\,\mu^2 \ .
\ee
Furthermore, if $\xi \notin \{0,\xi_c,\xi_*\}$, by repeating the procedure from above, we get that the Ricci scalar $R$ has to be a very specific constant,
\be\label{eq:RLamdmu}
R = \frac{2\xi(2\xi - 1)\Lambda - (4\xi - 1)^2 \mu^2}{16\xi^2(\xi - \xi_*)} \ ,
\ee
on the set $S - Z$. We shall exploit this constraint in the section 6. While it is straightforward to generalize the relation (\ref{eq:RLamdmu}) to the case of the general potential $V$, the result is pretty cumbersome and, more to the point, it does not seem to lead to a useful constraint on the Ricci scalar.

\medskip

Finally, even more ambitious line of generalization is to look ``beyond Einstein'', with the tensor $E_{ab}$ which satisfies the orthogonal-transitive condition (\ref{eq:ot1}). If we assume that $V(\phi) = 0$ and $\xi \notin \{0,\xi_c\}$, by repeating the same procedure as above we get
\be
T_{ab}^{(\phi)} K^a \nabla^b \phi = \left( \mathscr{E} + \frac{1}{2}\,(8\xi - 1)R \xi\phi^2 \right) \Lie_K \phi \ ,
\ee
where
\be
\mathscr{E} \equiv \frac{g^{ab} E_{ab}}{32\pi(D-1)(\xi - \xi_c)} \ ,
\ee
so that the equation (\ref{eq:EOMKnabphi}) on the set $S$ implies
\be\label{eq:ERgamma}
\mathscr{E} + \left( \frac{1}{2}\,(8\xi - 1)R + \gamma_{\mathrm{G}} \right) \xi \phi^2 - \frac{1}{8\pi}\,\gamma_{\mathrm{E}} = 0 \ .
\ee
We can extract $\gamma_G$ by taking the Lie derivative $\Lie_K$ of the last equation,
\be
\gamma_G = \frac{1}{2}\,(1 - 8\xi)R \ ,
\ee
and insert it back into the equation (\ref{eq:ERgamma}) to reduce it to
\be
g^{ab}E_{ab} = 4(D-1)(\xi - \xi_c) \gamma_E \ .
\ee
At this point we see the limitations of the procedure introduced above: The constraint we have established on the trace of the tensor $E_{ab}$ is still pretty implicit, as it does not give a clear relation between the tensor $E_{ab}$ and the parameters of the model that would need to be satisfied on the set $S$.

\medskip

\section{Black hole hair via symmetry noninheritance}

All previous discussion is general in a sense that we only look at the local properties of the spacetime. One particularly interesting application of the symmetry inheritance properties is in the context of black hole spacetimes and associated no-hair theorems. Ever since the early results on no-hair theorems, most of the activity in this field of research can be described as an exchange of challenges, the quest to find as general as possible constraints on the allowed form of the black hole hair on one side, and the quest to find the way to circumvent those constraints and provide hairy black hole solutions on the other. There is a compelling body of evidence, based on detailed numerical \cite{HR14,HR15} and analytic \cite{CSR15} analyses, that the complex scalar hair can circumvent the no-hair theorems via symmetry noninheritance. On the other hand, there are theorems \cite{ISm15,ISm17} which pretty much exclude the existence of the nontrivial symmetry noninheriting \emph{minimally coupled} real scalar hair in the presence of Killing horizons. Still, this leaves the possibility that \emph{nonminimally coupled} real scalar field might provide symmetry noninheriting black hole hair. 

\medskip

What is the current status of the no-hair theorems for the nonminimally coupled real scalar black hole hair? In the absence of the scalar self-interaction, that is when the potential is identically zero $V(\phi) = 0$, there are several lines of generalization \cite{XZ91,Saa96a,Saa96b,SZ98,Winst05,HW09} (see also discussion about the $(2+1)$-dimensional case in \cite{ABGMPS00}), all of which assume that the spacetime is spherically symmetric, static (at least in the domain of outer communications) and the scalar field $\phi$ inherits all symmetries (i.e.~it depends only on the radial coordinate, $\phi = \phi(r)$). Cases with nonvanishing potential $V$ are covered by the series of results in \cite{MB96} (for the spherically symmetric spacetimes, nonnegative potential $V \ge 0$ and $\xi < 0$ or $\xi \ge 1/2$) and \cite{AB02} (for the static spacetimes, quartic potential $V = \lambda \phi^4$ and any real $\xi$), both of which rely on the assumption about the symmetry inheritance.

\medskip

For any no-hair theorem which assumes the symmetry inheritance (and most of them do so without any justification) constraints presented in the sections 3 and 4 can either

\begin{itemize}
\item[(a)] close the gap is a sense that one cannot evade the no-hair theorem result on account of breaking of the symmetry inheritance, or

\item[(b)] point to the gap which is still open (since the constraints presented here do not guarantee the symmetry inheritance under the conditions considered by the given no-hair theorems) and which indicates the possibility of a novel symmetry noninheriting black hole hair.
\end{itemize}

To summarize, in the conformal case ($\xi = \xi_c$) there can be no symmetry noninheriting black hole hair unless $V = 0$ or $V = V_c$. Note that the latter two cases are exactly those which were (based on the principle of conformal invariance) chosen for the no-hair analyses. Furthermore, in the non-conformal case ($\xi \ne \xi_c$), situation is much more intricate, as explained by the theorem \ref{tm:zeroV} and the discussion below. For example, if the potential $V$ is vanishing the assumption on symmetry inheritance is unwarranted unless $R \ne 0$.

Of course, one might object that the gaps left open by the constraints above (e.g.~$V = V_c$ in the theorem \ref{tm:confxi} or $R = 0$ in the theorem \ref{tm:zeroV}) are just methodological artefacts. However, examples in the next section will demonstrate that this is not so. We turn now to the phenomenon of stealth fields, a rich source of spacetimes with symmetry noninheriting scalar fields.

\medskip

\section{Stealth scalar fields}

Although by Wheeler's maxim ``matter tells spacetime how to curve'', the effect of the matter on the spacetime can sometimes be seemingly invisible. A decade ago it was noticed \cite{ABMTZ05} that some exact solutions of the \emph{vacuum} Einstein's equations are simultaneously exact solutions of the Einstein-Klein-Gordon equations with non\-mini\-mally coupled scalar fields. More concretely, any nontrivial scalar field configuration that is a solution of the field equations (\ref{eq:EOM})--(\ref{eq:KG}), such that $T_{ab}^{(\phi)} + 2f(\phi)G_{ab} = 0$ and, consequently, $E_{ab} = 0$, was dubbed \emph{stealth} scalar field. It is worth noting that the minimally coupled scalar fields don't allow stealth configurations. Namely, if $f = 0$ and $T_{ab}^{(\phi)} = 0$, then the equations $g^{ab} T_{ab}^{(\phi)} = 0$ and $g^{ac} g^{bd} T_{ab}^{(\phi)} T_{cd}^{(\phi)} = 0$ imply that for any integer $D \ge 2$ we necessarily have $X = 0$ and $V(\phi) = 0$, from which it follows that $\nab{a}\phi \nab{b}\phi = 0$. In other words, in this case the field $\phi$ is just a constant.

\medskip

The analyses of the spacetimes ``dressed'' with the stealth scalar fields were done for the Minkowski spacetime \cite{ABMTZ05}, the BTZ black hole \cite{ABMZ06} and various cosmological spacetimes \cite{BJJ08,AGRT13,CCH16}. The analysis of the linear stability of the stealth scalar fields with respect to small tensor perturbations in \cite{FM10} has found that some regions of the parameter space correspond to stable configurations, making them viable astrophysical models. Since in any of these examples the background spacetime possesses a number of isometries and the stealth scalar field is, at least in the gravitational field equation, essentially decoupled from gravity, these solutions are plentiful source of symmetry noninheriting configurations. 

\medskip

Suppose we want to dress an exact $D \ge 3$ dimensional solution of the Einstein's field equation, $E_{ab} = G_{ab} + \Lambda g_{ab} = 0$, with a stealth scalar field. This amounts to finding a scalar field $\phi$, such that 
\be
T_{ab}^{(\phi)} = \xi \Lambda \phi^2 g_{ab} \qquad \textrm{and} \qquad \Box\phi = V'(\phi) + \frac{2D\Lambda}{D-2}\,\xi\phi
\ee
for some potential $V$ and parameter $\xi$. Here we may use similar tricks as in the previous section. The conformal case $\xi = \xi_c$ is basically covered by the discussion in the subsection 3.1. If $\xi \ne \xi_c$ then we can extract $X$ from the trace $T = D \xi \Lambda \phi^2$ and plug it back to the contraction 
\be
T_{ab}^{(\phi)} K^a \nabla^b\phi = \xi \Lambda \phi^2 \Lie_K \phi \ . 
\ee
The result is the equation 
\be
W_\xi(\phi) \Lie_K \phi = 0 \ ,
\ee
where we have introduced an auxiliary function
\begin{eqnarray}
W_\xi(\phi) \equiv & 2\xi^2 \phi^2 V''(\phi) + 3\xi(2\xi - 1) \phi V'(\phi) + (1 - 2\xi) V(\phi) + \nonumber\\
 & + \frac{\xi(\xi - \xi_c)(1 + (4\xi - 1)D)}{(D-1)\xi_c}\,\Lambda\phi^2 \label{eq:Wxi} \ .
\end{eqnarray}
The equation above implies that on the set $S$ we necessarily have $W_\xi(\phi) = 0$. Now, it is again interesting to see for which potentials $V$ the function $W_\xi$ identically vanishes, allowing the possibility of a ``substantially large'' set $S$. We have the following cases:

\begin{itemize}
\item[(a)] if $\xi \in \rr - \{0,\xi_c,1/4\}$ then
\be
V(\phi) = A\phi^{\frac{1}{2\xi}} + B\phi^{\frac{1-2\xi}{\xi}} - \frac{4\xi(\xi-\xi_c)(1 + (4\xi - 1)D)}{(D-2)(4\xi - 1)^2}\,\Lambda\phi^2 \label{eq:V1} \ ,
\ee

\item[(b)] if $\xi = 1/4$ then 
\be
V(\phi) = \left( A + B \ln\phi - \frac{\Lambda}{(D-1)(D-2)}\,(\ln\phi)^2 \right) \phi^2 \label{eq:V2} \ ,
\ee
\end{itemize}

\noindent
where $A$ and $B$ are some real constants. Let us now look at some physically relevant examples.

\medskip

\emph{Minkowski spacetime}. The $D$-dimensional Minkowski spacetime $(\rr^D,\eta_{ab})$ is a maximally symmetric space, so any stealth field configuration on top of it will easily break the symmetry inheritance. If we look at the case of vanishing potential $V$, the Minkowski spacetime automatically satisfies one of the constraints given by theorem \ref{tm:zeroV} (as its Riemann tensor is identically zero), so we need to find a scalar field with $X = 0$ (at least for the nonconformal case, $\xi \ne \xi_c$). Take any constant (nonvanishing) null vector field $\ell^a$ and $\phi = \phi(u)$, where $u = \eta_{\mu\nu} \ell^\mu x^\nu$. Then the Klein-Gordon equation is automatically satisfied,
\be
\Box\phi(u) = \eta_{\mu\nu} \ell^\mu \ell^\nu \phi''(u) = 0
\ee
and the kinetic term vanishes,
\be
-2X = \eta_{\mu\nu} \ell^\mu \ell^\nu (\phi'(u))^2 = 0 \ .
\ee
The gravitational field equation reduces to the nonlinear differential equation
\be
2\xi\,\phi(u) \phi''(u) + (2\xi - 1) (\phi'(u))^2 = 0 \ .
\ee
We have the following subcases

\begin{itemize}
\item[(a)] $\xi \in \rr - \{0,1/4\}$
\be
\phi(u) = \left( \alpha u + \beta \right)^{\frac{2\xi}{4\xi-1}} \ ,
\ee

\item[(b)] $\xi = 1/4$
\be
\phi(u) = \alpha e^{\beta u} \ ,
\ee
\end{itemize}

\noindent
where $\alpha$ and $\beta$ are some real constants. In both of these classes of solutions the symmetry inheritance is manifestly broken. The systematic study of dressing of the Minkowski spacetime with the nonminimally coupled real scalar field has been done in \cite{ABMTZ05}, where the authors have constructed the general form of the potential $V$ which allows such stealth solutions. These potentials are exactly those to which the potentials (\ref{eq:V1})--(\ref{eq:V2}) reduce in the $\Lambda = 0$ case, as well equal to the potential (\ref{eq:Vc}) in the conformal case. Note, however, that our conclusions hold for any exact solution of the Einstein's gravitational field equation which admits an isometry that is broken by the stealth scalar field. Also, if $V = V_{\mathrm{mass}}$ with $\mu \ne 0$, then the constraint (\ref{eq:RLamdmu}), with $R = 0$ and $\Lambda = 0$, implies that for $\xi \notin \{0,\xi_*\}$ the coupling constant must be $\xi = 1/4$ (with an exception of the case $D = 3$, for which $\xi_* = 1/4$), and an example of such a stealth field can be also found in \cite{ABMTZ05}.

\medskip

\emph{Cosmological spacetimes}. Analogous classes of stealth scalar fields can be con\-struc\-ted on top of the homogeneous isotropic universe. These have been studied in \cite{AGRT13}, using conformal coupling $\xi = \xi_c$ and the potential of the form (\ref{eq:Vc}). Apart from the form of the potentials given by (\ref{eq:V1})--(\ref{eq:V2}), here we may comment on the case when $V = V_{\mathrm{mass}}$ and the cosmological solution admits a hypersurface orthogonal Killing vector field (e.g.~FRW solutions with the flat spatial geometry). If we combine the equation (\ref{eq:RLamdmu}) with $R = 2D\Lambda/(D-2)$, we get a constraint on the parameter $\xi$ in a form of the cubic equation,
\be
\frac{2D}{D-2}\,\Lambda = \frac{2\xi(2\xi - 1)\Lambda - (4\xi - 1)^2 \mu^2}{16\xi^2(\xi - \xi_*)} \ .
\ee
For example, if $\mu = 0$ and $\Lambda \ne 0$, then this equation implies that we must have either $\xi = \xi_c$ or $\xi = (D-1)/4D$. We note in passing that the authors of \cite{MM12,BGH13a,BGH13b} consider cosmological solutions dressed with stealth, but symmetry inheriting scalar fields.

\medskip

\emph{Black hole spacetimes}. Finally, it is tempting to ask if it is possible to dress up a black hole with a stealth scalar field. As was explained in the previous section, all the gaps left by the symmetry inheritance constraints indicate the possible way to evade the known no-hair theorems. Indeed, examples of stealth scalar fields on the 3-dimensional BTZ black hole have been found in \cite{ABMZ06}. Here we have time dependent stealth fields (for various potentials $V$) on top of the stationary spacetime. Still, physically more interesting cases are those of the $1+3$ spacetime dimensions. To our knowledge, no such solutions, of 4-dimensional black holes with the symmetry noninheriting stealth scalar hair, have been found (or proven not to exist by some novel no-hair theorem). Let us look, for example, at the Schwarzschild spacetime in which the exterior of the black hole is covered with the coordinate system $\{t,r,\theta,\varphi\}$. If we assume that $\dd_\theta \phi = 0$, then the equation $T^{(\phi)}_{\theta\varphi} = 0$ implies that in fact $\dd_\varphi \phi = 0$, so we have $\phi = \phi(t,r)$. Furthermore, if we integrate the equation $\tensor{T}{^r_r} - \tensor{T}{^\theta_\theta} = 0$ and then insert the result into the equation $T^{(\phi)}_{tr} = 0$, we get that $\dd_t \phi = 0$, thus $\phi = \phi(r)$ and this is a symmetry inheriting field. This, however, is not enough to prove that such a hair does not exist due to the remaining case when $\dd_\theta \phi \ne 0$.

\medskip

\section{Final remarks}

The methods of approach to the problem of symmetry inheritance of the nonminimally coupled fields and the various classification schemes presented in this paper should be taken as the first step towards the more comprehensive understanding of the breaking of the symmetry inheritance. Several obvious questions remain open for the future work. First of all, our method has a ``blind spot'' in the $\xi = \xi_*$ case. A further study is needed for the other cases of the function $f$, which defines the nonminimal coupling, as well for the more general potentials $V$ when the coupling is not conformally invariant. Also, our method does not seem to give useful constraints with some more complicated cases of the nonminimal coupling, such is the one in the EGBd model \cite{MS93,KKMR16} (where the scalar field is coupled to the Gauss-Bonnet Lagrangian term) or in the generalized Brans-Dicke theories \cite{DeFT10BD,RFVM14}, thus a different approach is needed. 

Ramifications of the symmetry inheritance results on the no-hair theorems have been discussed in the section 5 and at the end of the section 6. For example, in the conformal case (with $\xi = \xi_c$ and $V = V_c$) stealth field configurations demonstrate the existence of the symmetry noninheriting solutions and the associated gap in the no-hair theorems (which assume that the scalar field inherits the symmetries). The most important question left open here is whether there might exist a 4-dimensional black hole solution with nonminimally coupled, symmetry noninheriting scalar hair. The classification of the symmetry noninheriting black hole hair \cite{ISm15,ISm17} is still in its infancy, and here one might hope for some further refinements of the no-hair theorems and possibly novel examples of such hair.

\ack
We would like to thank Edgardo Franzin for the series of valuable comments on the first draft of the paper. This research has been supported by the Croatian Science Foundation under the project No.~8946.

\section*{References}

\bibliographystyle{iopnum}
\bibliography{nonminscal}

\providecommand{\newblock}{}
\begin{thebibliography}{10}
\expandafter\ifx\csname url\endcsname\relax
  \def\url#1{{\tt #1}}\fi
\expandafter\ifx\csname urlprefix\endcsname\relax\def\urlprefix{URL }\fi
\providecommand{\eprint}[2][]{\href{http://arxiv.org/abs/#2}{#2}}

\bibitem{Woo73a}
Woolley M 1973 {\em Comm. Math. Phys.\/}
  \href{http://dx.doi.org/10.1007/BF01645590}{{\bf {\bf 31}} 75--81}

\bibitem{Woo73b}
Woolley M 1973 {\em Comm. Math. Phys.\/}
  \href{http://dx.doi.org/10.1007/BF01645625}{{\bf {\bf 33}} 135--144}

\bibitem{MW75}
Michalski H and Wainwright J 1975 {\em Gen. Relativ. Gravit.\/}
  \href{http://dx.doi.org/10.1007/BF00751574}{{\bf {\bf 6}} 289--318}

\bibitem{Coll75}
Coll B 1975 {\em C. R. Acad. Sci. (Paris) A\/} {\bf {\bf 280}} 1773--1776
  \urlprefix\url{http://gallica.bnf.fr/ark:/12148/bpt6k62167964/f357.item.r=1773}

\bibitem{RT75}
Ray J~R and Thompson E~L 1975 {\em J. Math. Phys.\/}
  \href{http://dx.doi.org/10.1063/1.522548}{{\bf {\bf 16}} 345--346}

\bibitem{WY76a}
Wainwright J and Yaremovicz P~E~A 1976 {\em Gen. Relativ. Gravit.\/}
  \href{http://dx.doi.org/10.1007/BF00771105}{{\bf {\bf 7}} 345--359}
  \urlprefix\url{https://link.springer.com/article/10.1007/BF00771105}

\bibitem{WY76b}
Wainwright J and Yaremovicz P~A~E 1976 {\em Gen. Relativ. Gravit.\/}
  \href{http://dx.doi.org/10.1007/BF00763408}{{\bf {\bf 7}} 595--608}

\bibitem{Hoen78}
Hoenselaers C 1978 {\em Prog. Theor. Phys.\/}
  \href{http://dx.doi.org/10.1143/PTP.59.1518}{{\bf {\bf 59}} 1518--1521}

\bibitem{ISm15}
Smoli{\'c} I 2015 {\em Class. Quantum Grav.\/}
  \href{http://dx.doi.org/10.1088/0264-9381/32/14/145010}{{\bf {\bf 32}}
  145010} ({\em Preprint} \eprint{1501.04967})

\bibitem{CDPS16}
Cvitan M, Dominis~Prester P and Smoli{\'c} I 2016 {\em Class. Quantum Grav.\/}
  \href{http://dx.doi.org/10.1088/0264-9381/33/7/077001}{{\bf {\bf 33}} 077001}
  \href{https://cqgplus.com/2016/03/23/perfect-accordance-of-the-gravitational-and-the-electromagnetic-field-in-3d/}{CQG+}
  ({\em Preprint} \eprint{1508.03343})

\bibitem{ISm17}
Smoli{\'c} I 2017 {\em Phys. Rev.\/}
  \href{http://dx.doi.org/10.1103/PhysRevD.95.024016}{{\bf D {\bf 95}} 024016}
  ({\em Preprint} \eprint{1609.04013})

\bibitem{BGS17}
Barja{\v s}i{\'c} I, Gulin L and Smoli{\'c} I 2017 {\em Phys. Rev.\/}
  \href{http://dx.doi.org/10.1103/PhysRevD.95.124037}{{\bf D {\bf 95}} 124037}
  ({\em Preprint} \eprint{1705.00628})

\bibitem{BD}
Birrell N~D and Davies P~C~W 1982 {\em Quantum fields in curved space\/}
  (Cambridge Cambridgeshire New York: Cambridge University Press) ISBN
  978-0521233859

\bibitem{CCJ70}
Callan Jr C~G, Coleman S~R and Jackiw R 1970 {\em Annals Phys.\/}
  \href{http://dx.doi.org/10.1016/0003-4916(70)90394-5}{{\bf {\bf 59}} 42--73}

\bibitem{FM87}
Futamase T and Maeda K 1989 {\em Phys. Rev.\/}
  \href{http://dx.doi.org/10.1103/PhysRevD.39.399}{{\bf D39} 399--404}

\bibitem{Faraoni96}
Faraoni V 1996 {\em Phys. Rev.\/}
  \href{http://dx.doi.org/10.1103/PhysRevD.53.6813}{{\bf D53} 6813--6821} ({\em
  Preprint} \eprint{astro-ph/9602111})

\bibitem{Saa96a}
Saa A 1996 {\em J. Math. Phys.\/}
  \href{http://dx.doi.org/10.1063/1.531513}{{\bf {\bf 37}} 2346--2351} ({\em
  Preprint} \eprint{gr-qc/9601021})

\bibitem{Saa96b}
Saa A 1996 {\em Phys. Rev.\/}
  \href{http://dx.doi.org/10.1103/PhysRevD.53.7377}{{\bf D {\bf 53}}
  7377--7380} ({\em Preprint} \eprint{gr-qc/9602061})

\bibitem{AB02}
Ay{\'o}n-Beato E 2002 {\em Class. Quantum Grav.\/}
  \href{http://dx.doi.org/10.1088/0264-9381/19/21/311}{{\bf {\bf 19}}
  5465--5472} ({\em Preprint} \eprint{gr-qc/0212050})

\bibitem{Winst05}
Winstanley E 2005 {\em Class. Quantum Grav.\/}
  \href{http://dx.doi.org/10.1088/0264-9381/22/11/020}{{\bf {\bf 22}}
  2233--2248} ({\em Preprint} \eprint{gr-qc/0501096})

\bibitem{HW09}
Hosler D and Winstanley E 2009 {\em Phys. Rev.\/}
  \href{http://dx.doi.org/10.1103/PhysRevD.80.104010}{{\bf D {\bf 80}} 104010}
  ({\em Preprint} \eprint{0907.1487})

\bibitem{Beken72}
Bekenstein J~D 1972 {\em Phys. Rev. Lett.\/}
  \href{http://dx.doi.org/10.1103/PhysRevLett.28.452}{{\bf {\bf 28}} 452--455}

\bibitem{BekenStaticHair}
Bekenstein J~D 1972 {\em Phys. Rev.\/}
  \href{http://dx.doi.org/10.1103/PhysRevD.5.1239}{{\bf D {\bf 5}} 1239--1246}

\bibitem{BekenStationaryHair}
Bekenstein J~D 1972 {\em Phys. Rev.\/}
  \href{http://dx.doi.org/10.1103/PhysRevD.5.2403}{{\bf D {\bf 5}} 2403--2412}

\bibitem{XZ91}
Xanthopoulos B~C and Zannias T 1991 {\em J. Math. Phys.\/}
  \href{http://dx.doi.org/10.1063/1.529253}{{\bf {\bf 32}} 1875--1880}

\bibitem{Beken95}
Bekenstein J~D 1995 {\em Phys. Rev.\/}
  \href{http://dx.doi.org/10.1103/PhysRevD.51.R6608}{{\bf D {\bf 51}}
  R6608--R6611}

\bibitem{Sud95}
Sudarsky D 1995 {\em Class. Quantum Grav.\/}
  \href{http://dx.doi.org/10.1088/0264-9381/12/2/023}{{\bf {\bf 12}} 579--584}

\bibitem{Zann95}
Zannias T 1995 {\em J. Math. Phys.\/}
  \href{http://dx.doi.org/10.1063/1.531201}{{\bf {\bf 36}} 6970--6980} ({\em
  Preprint} \eprint{gr-qc/9409030})

\bibitem{SZ98}
Sudarsky D and Zannias T 1998 {\em Phys. Rev.\/}
  \href{http://dx.doi.org/10.1103/PhysRevD.58.087502}{{\bf D {\bf 58}} 087502}
  ({\em Preprint} \eprint{gr-qc/9712083})

\bibitem{BL07}
Bhattacharya S and Lahiri A 2007 {\em Phys. Rev. Lett.\/}
  \href{http://dx.doi.org/10.1103/PhysRevLett.99.201101}{{\bf {\bf 99}} 201101}
  ({\em Preprint} \eprint{gr-qc/0702006})

\bibitem{BL11}
Bhattacharya S and Lahiri A 2011 {\em Phys. Rev.\/}
  \href{http://dx.doi.org/10.1103/PhysRevD.83.124017}{{\bf D {\bf 83}} 124017}
  ({\em Preprint} \eprint{1102.0053})

\bibitem{SS14}
Sk{\'a}kala J and Shankaranarayanan S 2014 {\em Class. Quantum Grav.\/}
  \href{http://dx.doi.org/10.1088/0264-9381/31/17/175005}{{\bf {\bf 31}}
  175005} ({\em Preprint} \eprint{1402.6166})

\bibitem{Hod17}
Hod S 2017 {\em Physics Letters B\/}
  \href{http://dx.doi.org/10.1016/j.physletb.2017.06.005}{{\bf {\bf 771}}
  521--523}

\bibitem{Heusler}
Heusler M 1996 {\em {Black Hole Uniqueness Theorems}\/} (Cambridge New York:
  Cambridge University Press) ISBN 9780521567350

\bibitem{BJLP14}
Broderick A~E, Johannsen T, Loeb A and Psaltis D 2014 {\em Astrophys. J.\/}
  \href{http://dx.doi.org/10.1088/0004-637X/784/1/7}{{\bf {\bf 784}} 7} ({\em
  Preprint} \eprint{1311.5564})

\bibitem{CHRR15}
Cunha P~V~P, Herdeiro C~A~R, Radu E and R{\'u}narsson H~F 2015 {\em Phys. Rev.
  Lett.\/} \href{http://dx.doi.org/10.1103/PhysRevLett.115.211102}{{\bf {\bf
  115}} 211102} ({\em Preprint} \eprint{1509.00021})

\bibitem{Test16}
Johannsen T, Broderick A~E, Plewa P~M, Chatzopoulos S, Doeleman S~S, Eisenhauer
  F, Fish V~L, Genzel R, Gerhard O and Johnson M~D 2016 {\em Phys. Rev.
  Lett.\/} \href{http://dx.doi.org/10.1103/PhysRevLett.116.031101}{{\bf {\bf
  116}} 031101} ({\em Preprint} \eprint{1512.02640})

\bibitem{CG16}
Cardoso V and Gualtieri L 2016 {\em Class. Quantum Grav.\/}
  \href{http://dx.doi.org/10.1088/0264-9381/33/17/174001}{{\bf {\bf 33}}
  174001} ({\em Preprint} \eprint{1607.03133})

\bibitem{J16}
Johannsen T 2016 {\em Class. Quantum Grav.\/}
  \href{http://dx.doi.org/10.1088/0264-9381/33/12/124001}{{\bf 33} 124001}
  ({\em Preprint} \eprint{1602.07694})

\bibitem{Cast17}
Castelvecchi D 2017 {\em Nature\/}
  \href{http://dx.doi.org/10.1038/543478a}{{\bf {\bf 543}} 478--480}

\bibitem{BCW06}
Berti E, Cardoso V and Will C~M 2006 {\em Phys. Rev.\/}
  \href{http://dx.doi.org/10.1103/PhysRevD.73.064030}{{\bf D {\bf 73}} 064030}
  ({\em Preprint} \eprint{gr-qc/0512160})

\bibitem{GVS12}
Gossan S, Veitch J and Sathyaprakash B~S 2012 {\em Phys. Rev.\/}
  \href{http://dx.doi.org/10.1103/PhysRevD.85.124056}{{\bf D {\bf 85}} 124056}
  ({\em Preprint} \eprint{1111.5819})

\bibitem{Berti15}
Berti E {\em et~al.\/} 2015 {\em Class. Quantum Grav.\/}
  \href{http://dx.doi.org/10.1088/0264-9381/32/24/243001}{{\bf {\bf 32}}
  243001} ({\em Preprint} \eprint{1501.07274})

\bibitem{LIGO16}
Abbott B~P {\em et~al.\/} (Virgo, LIGO Scientific) 2016 {\em Phys. Rev.
  Lett.\/} \href{http://dx.doi.org/10.1103/PhysRevLett.116.061102}{{\bf {\bf
  116}} 061102} ({\em Preprint} \eprint{1602.03837})

\bibitem{CFP16}
Cardoso V, Franzin E and Pani P 2016 {\em Phys. Rev. Lett.\/}
  \href{http://dx.doi.org/10.1103/PhysRevLett.116.171101}{{\bf {\bf 116}}
  171101} ({\em Preprint} \eprint{1602.07309})

\bibitem{CHMPP16}
Cardoso V, Hopper S, Macedo C~F~B, Palenzuela C and Pani P 2016 {\em Phys.
  Rev.\/} \href{http://dx.doi.org/10.1103/PhysRevD.94.084031}{{\bf {\bf D94}}
  084031} ({\em Preprint} \eprint{1608.08637})

\bibitem{CFMPR17}
Cardoso V, Franzin E, Maselli A, Pani P and Raposo G 2017 {\em Phys. Rev.\/}
  \href{http://dx.doi.org/10.1103/PhysRevD.95.084014}{{\bf {\bf D95}} 084014}
  ({\em Preprint} \eprint{1701.01116})

\bibitem{CP17}
Cardoso V and Pani P 2017 {\em Nat. Astron.\/}
  \href{http://dx.doi.org/10.1038/s41550-017-0225-y}{{\bf {\bf 1}} 586--591}
  ({\em Preprint} \eprint{1709.01525})

\bibitem{BCRG17}
Barcel{\'o} C, Carballo-Rubio R and Garay L~J 2017 {\em JHEP\/}
  \href{http://dx.doi.org/10.1007/JHEP05(2017)054}{{\bf {\bf 05}} 054} ({\em
  Preprint} \eprint{1701.09156})

\bibitem{Wald}
Wald R 1984 {\em {General Relativity}\/} (Chicago: University of Chicago Press)
  ISBN 0226870332

\bibitem{Hryc17}
Hrycyna O 2017 {\em Phys. Lett. B\/}
  \href{http://dx.doi.org/10.1016/j.physletb.2017.02.062}{{\bf {\bf 768}}
  218--227} ({\em Preprint} \eprint{1511.08736})

\bibitem{Love71}
Lovelock D 1971 {\em J. Math. Phys.\/}
  \href{http://dx.doi.org/10.1063/1.1665613}{{\bf {\bf 12}} 498--501}

\bibitem{NO06}
Nojiri S and Odintsov S~D 2006 {\em eConf\/}
  \href{http://dx.doi.org/10.1142/S0219887807001928}{{\bf C0602061} 06} [Int.
  J. Geom. Meth. Mod. Phys. 4, 115 (2007)] ({\em Preprint}
  \eprint{hep-th/0601213})

\bibitem{NO10}
Nojiri S and Odintsov S~D 2011 {\em Phys. Rept.\/}
  \href{http://dx.doi.org/10.1016/j.physrep.2011.04.001}{{\bf {\bf 505}}
  59--144} ({\em Preprint} \eprint{1011.0544})

\bibitem{SF10}
Sotiriou T~P and Faraoni V 2010 {\em Rev. Mod. Phys.\/}
  \href{http://dx.doi.org/10.1103/RevModPhys.82.451}{{\bf {\bf 82}} 451--497}
  ({\em Preprint} \eprint{0805.1726})

\bibitem{DeFT10}
De~Felice A and Tsujikawa S 2010 {\em Living Rev. Rel.\/}
  \href{http://dx.doi.org/10.12942/lrr-2010-3}{{\bf {\bf 13}} 3} ({\em
  Preprint} \eprint{1002.4928})

\bibitem{HR14}
Herdeiro C~A~R and Radu E 2014 {\em Phys. Rev. Lett.\/}
  \href{http://dx.doi.org/10.1103/PhysRevLett.112.221101}{{\bf {\bf 112}}
  221101} ({\em Preprint} \eprint{1403.2757})

\bibitem{HR15}
Herdeiro C~A~R and Radu E 2015 {\em Int. J. Mod. Phys.\/}
  \href{http://dx.doi.org/10.1142/S0218271815420146}{{\bf D {\bf 24}} 1542014}
  ({\em Preprint} \eprint{1504.08209})

\bibitem{CSR15}
Chodosh O and Shlapentokh-Rothman Y 2017 {\em Commun. Math. Phys.\/}
  \href{http://dx.doi.org/10.1007/s00220-017-2998-3}{{\bf {\bf 356}}
  1155--1250} ({\em Preprint} \eprint{1510.08025})

\bibitem{ABGMPS00}
Ay{\'o}n-Beato E, Garc{\'i}a A, Mac{\'i}as A and P{\'e}rez-S{\'a}nchez J~M 2000
  {\em Phys. Lett.\/}
  \href{http://dx.doi.org/10.1016/S0370-2693(00)01241-7}{{\bf B495} 164--168}
  ({\em Preprint} \eprint{gr-qc/0101079})

\bibitem{MB96}
Mayo A~E and Bekenstein J~D 1996 {\em Phys. Rev.\/}
  \href{http://dx.doi.org/10.1103/PhysRevD.54.5059}{{\bf D54} 5059--5069} ({\em
  Preprint} \eprint{gr-qc/9602057})

\bibitem{ABMTZ05}
Ay{\'o}n-Beato E, Mart{\'i}nez C, Troncoso R and Zanelli J 2005 {\em Phys.
  Rev.\/} \href{http://dx.doi.org/10.1103/PhysRevD.71.104037}{{\bf D {\bf 71}}
  104037} ({\em Preprint} \eprint{hep-th/0505086})

\bibitem{ABMZ06}
Ay{\'o}n-Beato E, Mart{\'i}nez C and Zanelli J 2006 {\em Gen.~Rel.~Grav.\/}
  \href{http://dx.doi.org/10.1007/s10714-005-0213-x}{{\bf {\bf 38}} 145--152}
  ({\em Preprint} \eprint{hep-th/0403228})

\bibitem{BJJ08}
Banerjee N, Jain R~K and Jatkar D~P 2008 {\em Gen.~Rel.~Grav.\/}
  \href{http://dx.doi.org/10.1007/s10714-007-0516-1}{{\bf 40} 93--105} ({\em
  Preprint} \eprint{hep-th/0610109})

\bibitem{AGRT13}
Ay{\'o}n-Beato E, Garc{\'i}a A~A, Ram{\'i}rez-Baca P~I and Terrero-Escalante
  C~A 2013 {\em Phys. Rev.\/}
  \href{http://dx.doi.org/10.1103/PhysRevD.88.063523}{{\bf D {\bf 88}} 063523}
  ({\em Preprint} \eprint{1307.6534})

\bibitem{CCH16}
Campuzano C, C{\'a}rdenas V~H and Herrera R 2016 {\em Eur. Phys. J.\/}
  \href{http://dx.doi.org/10.1140/epjc/s10052-016-4546-2}{{\bf C {\bf 76}} 698}
  ({\em Preprint} \eprint{1611.08433})

\bibitem{FM10}
Faraoni V and Moreno A~F~Z 2010 {\em Phys. Rev.\/}
  \href{http://dx.doi.org/10.1103/PhysRevD.81.124050}{{\bf D {\bf 81}} 124050}
  ({\em Preprint} \eprint{1006.1936})

\bibitem{MM12}
Maeda H and Maeda K 2012 {\em Phys. Rev.\/}
  \href{http://dx.doi.org/10.1103/PhysRevD.86.124045}{{\bf D86} 124045} ({\em
  Preprint} \eprint{1208.5777})

\bibitem{BGH13a}
Bravo~Gaete M and Hassa{\"i}ne M 2013 {\em Phys. Rev.\/}
  \href{http://dx.doi.org/10.1103/PhysRevD.88.104011}{{\bf D {\bf 88}} 104011}
  ({\em Preprint} \eprint{1308.3076})

\bibitem{BGH13b}
Bravo~Gaete M and Hassa{\"i}ne M 2013 {\em JHEP\/}
  \href{http://dx.doi.org/10.1007/JHEP11(2013)177}{{\bf {\bf 11}} 177} ({\em
  Preprint} \eprint{1309.3338})

\bibitem{MS93}
Mignemi S and Stewart N~R 1993 {\em Phys. Rev.\/}
  \href{http://dx.doi.org/10.1103/PhysRevD.47.5259}{{\bf D {\bf 47}}
  5259--5269} ({\em Preprint} \eprint{hep-th/9212146})

\bibitem{KKMR16}
Kleihaus B, Kunz J, Mojica S and Radu E 2016 {\em Phys. Rev.\/}
  \href{http://dx.doi.org/10.1103/PhysRevD.93.044047}{{\bf D {\bf 93}} 044047}
  ({\em Preprint} \eprint{1511.05513})

\bibitem{DeFT10BD}
De~Felice A and Tsujikawa S 2010 {\em JCAP\/}
  \href{http://dx.doi.org/10.1088/1475-7516/2010/07/024}{{\bf {\bf 1007}} 024}
  ({\em Preprint} \eprint{1005.0868})

\bibitem{RFVM14}
Rasouli S~M~M, Farhoudi M and Vargas~Moniz P 2014 {\em Class. Quantum Grav.\/}
  \href{http://dx.doi.org/10.1088/0264-9381/31/11/115002}{{\bf {\bf 31}}
  115002} ({\em Preprint} \eprint{1405.0229})

\end{thebibliography}

\end{document}